\documentclass{article}

\usepackage[numbers,square]{natbib}
\usepackage{arxiv}
\usepackage[utf8]{inputenc} % allow utf-8 input
\usepackage[T1]{fontenc}    % use 8-bit T1 fonts
\usepackage{hyperref}       % hyperlinks
\usepackage{url}            % simple URL typesetting
\usepackage{booktabs}       % professional-quality tables
\usepackage{amsfonts}       % blackboard math symbols
\usepackage{nicefrac}       % compact symbols for 1/2, etc.
\usepackage{microtype}      % microtypography
\usepackage{graphicx}
\usepackage{subcaption}
\usepackage[hang,flushmargin]{footmisc}
\usepackage{enumitem}
\usepackage{xcolor}
\usepackage[section]{placeins}
\usepackage{longtable}
\usepackage{caption}
\usepackage{tabularx}
\usepackage{parnotes}

\title{
Identifying the Development and Application of Artificial Intelligence in Scientific Text\thanks{\noindent We thank Kevin Boyack, Daniel Chou, Teddy Collins, Dick Klavans, and Ilya Rahkovsky for their feedback and ideas on this work. We are grateful to the team at Elsevier for extended discussions about the methodological details of a related project, and sharing expert-curated keywords and labeled data. Zihe Yang led the replication of the Elsevier approach to identifying AI-relevant research. Neha Tiwari contributed the descriptive analysis of arXiv and conference-paper data, and assisted with model development. For replication materials, see  \url{https://github.com/georgetown-cset/ai-relevant-papers}.}
}

\author{
    James Dunham\\
    Center for Security and Emerging Technology\\
    Georgetown University\\
    \href{mailto:james.dunham@georgetown.edu}{\texttt{james.dunham@georgetown.edu}}\\
    \hspace{2em}
    \And Jennifer Melot\\
    Center for Security and Emerging Technology\\ 
    Georgetown University\\
    \href{mailto:jennifer.melot@georgetown.edu}{\texttt{jennifer.melot@georgetown.edu}}\\
    \hspace{2em}
    \And Dewey Murdick\\
    Center for Security and Emerging Technology\\ 
    Georgetown University\\
    \href{mailto:dewey.murdick@georgetown.edu}{\texttt{dewey.murdick@georgetown.edu}}
}

\begin{document}
\maketitle

\begin{abstract}
\noindent We describe a strategy for identifying the universe of research publications relevant to the application and development of artificial intelligence.
The approach leverages the arXiv corpus of scientific preprints, in which authors choose subject tags for their papers from a set defined by editors.
We compose a functional definition of AI relevance by learning these subjects from paper metadata, and then inferring the arXiv-subject labels of papers in larger corpora: Clarivate Web of Science, Digital Science Dimensions, and Microsoft Academic Graph.
This yields predictive classification $F_1$ scores between .75 and .86 for Natural Language Processing (\texttt{cs.CL}), Computer Vision (\texttt{cs.CV}), and Robotics (\texttt{cs.RO}).
For a single model that learns these and four other AI-relevant subjects (\texttt{cs.AI}, \texttt{cs.LG}, \texttt{stat.ML}, and \texttt{cs.MA}), we see precision of .83 and recall of .85.
We evaluate the out-of-domain performance of our classifiers against other sources of topic information and predictions from alternative methods.
We find that a supervised solution can generalize to identify publications that belong to the high-level fields of study represented on arXiv.
This offers a method for identifying AI-relevant publications that updates at the pace of research output, without reliance on subject-matter experts for query development or labeling.
\end{abstract}

\section{Overview}

Study of the applications and development of artificial intelligence faces a definitional problem: AI is a moving conceptual target, understood differently across researchers and observers of the field \citep{krafft_defining_2019}.
This presents a challenge for analysts and policy-makers \citep{tarraf_department_2019}.
The proliferation of reports on AI describe only partially overlapping domains \citep{elsevier_artificial_2018,raymond_perrault_ai_2019,allen_understanding_2019}, so their conclusions may be sensitive to the delineation of the field~\citep{zitt_meso-level_2015}.
We describe a strategy for addressing this and identifying a universe of AI-relevant scientific publications for use in bibliometric work.

The approach relies on the success of Cornell's arXiv project in attracting open-access preprints from subfields of computer science, physics, statistics, and other quantitative fields.\footnote{~\url{https://arxiv.org}.}
Authors and editors choose subject tags for these papers.
There are 39 subjects in computer science, including those we will consider relevant to AI: Artificial Intelligence, Computer Vision, Computation and Language (Natural Language Processing), Machine Learning, Multiagent Learning, and Robotics.\footnote{For the full taxonomy, see~\url{https://arxiv.org/category_taxonomy}.}
The arXiv labels offer a particular ground truth defined by the participation of an expert community.
Additionally, arXiv's implicit definition of subjects has the highly desirable characteristic of updating in real time, as opposed to less-favorable approaches that rely on keyword curation or annotation by subject-matter experts.
Those alternatives tend to require maintenance over time, and as we demonstrate, a query that subject-matter experts calibrate to retrieve AI-relevant publications in 2019 may struggle to surface those from 2010.

We are keenly aware that the subjects comprising AI research and applications are contestable.
Rather than argue for a single delineation, we offer an approach which requires only that an operational definition is composable from the subjects available to arXiv authors.
The sensitivity of all subsequent analysis to that choice of relevant subjects can be assessed through ablation. Researchers may also add or remove particular subjects as appropriate for their analyses.

We implement this approach by training SciBERT \citep{Beltagy2019Scibert} classifiers on arXiv metadata and subject labels.
Using the arXiv-trained models, we infer the subject relevance of papers in other corpora.
The premise of identifying AI-relevant publications in this way is that a model trained on arXiv data will successfully generalize to other sets of publication data, which may significantly differ in content and subject distribution.
This approach seems plausible when leveraging SciBERT's pre-training, but the risk of overfitting to arXiv and gaps in its coverage are concerns we address below with a series of results.

First, to assess performance within arXiv, we evaluate our models on a test set.
We observe $F_1$ scores between .75 and .86 for three subject-specific models, and .84 for a model trained on labels collapsed to indicate AI-relevance for papers in any of six AI-relevant subjects.
For comparison, we also assess a keyword-query solution and a keyword-learner hybrid developed for a recent bibliometric analysis of AI-relevant publications in Scopus \citep{elsevier_artificial_2018,raymond_perrault_ai_2019}.
Evaluation against arXiv labels yields $F_1$ scores of .55 and .59, respectively, for these methods.

We then report results from applying the models to scientific text in larger corpora: Clarivate Web of Science (WoS),\footnote{~\url{https://clarivate.com/webofsciencegroup}.} Digital Science Dimensions,\footnote{~\url{https://www.digital-science.com/products/dimensions/}.} and Microsoft Academic Graph (MAG).\footnote{~\url{https://www.microsoft.com/en-us/research/project/microsoft-academic-graph}.}
In the absence of ground-truth arXiv labels from these sources, we assess out-of-domain performance using other sources of topic information, by showing rates of predicted subject relevance in the fields of study defined by MAG.
We find that in the fields represented on arXiv, generalizing for inference in other corpora is feasible.
This offers a method for identifying AI-relevant publications that updates at the pace of research output, without reliance on subject-matter experts for query development or labeling.

\section{Development and applications of artificial intelligence}
\label{sec:ai}

Scientific text offers insight into the development of a field: its analysis can identify the organization of research communities; their breakthroughs or stagnation; and progress from basic research to applications \citep[e.g.,][]{small_identifying_2014, boyack_classification_2014}.
The obstacles to such inference are delineation of that field and the identification of emergent topics or technologies within it.\footnote{~For a discussion of precisely what constitutes emerging technology, see \citep{suominen2017exploring}.} In reference to biotech and nanotech in prior decades, \citeauthor{mogoutov_data_2007} write, ``Their content and dynamic are difficult to track at a time when they are struggling to define what they are, what they include and exclude, and how they organize and classify themselves internally'' \citep{mogoutov_data_2007}.
A related problem is identifying as-yet-unknown topics within a field, without the benefit of historical perspective.
Even in emergent areas, the distinction between ``legacy technologies'' and ``emerging technology'' may be incremental \citep{huang_systematic_2015}.

Recent analyses of AI research using query-based methods to delineate the field \citep{niu_global_2016, miyazaki_analyses_2018, rincon-patino2018exploring} have encountered these obstacles.
Grappling with the problem of query development in bibliometric work on nanotechnology resulted in principled methods for term curation and their evaluation \citep{mogoutov_data_2007, arora2013capturing, huang2011nanoscience, milanez_delineating_2016}, from which studies of AI could benefit.
Drawing from this literature, for example, \citeauthor{huang_systematic_2015} develop a method for retrieving ``big data'' research that expands from an initial set of terms across iterations of discovery, manual review, expert checks, and tuning for performance \citep{huang_systematic_2015}.

Other approaches to delineation depend on or begin with the identification of relevant journals \citep{gao_bibliometric_2019} or conferences \citep{martinez-plumed_facets_2018, shukla2019engineering}.
While appropriate for some analytic purposes, this method risks omitting relevant research in more general-audience venues or other disciplines, which may be a particularly acute problem for AI.

In review of the variety of methods for delineating the field of AI-relevant research, we note that beyond the methodological difficulties, the criteria for a system's intelligence vary by observer and over time.
In the typology developed by \citet{russell_artificial_2009}, definitions may emphasize behavior or reasoning, and evaluate it against human or rational standards.
In recent survey research \citep{krafft_defining_2019}, AI researchers tended to prefer definitions that emphasized the correctness of decisions and actions, but often disagreed on what satisfied these requirements.

Our own interest in high-quality analysis of AI and its security implications\footnote{\
    The Center for Security and Emerging Technology (CSET) studies the security impacts of emerging technologies and delivers nonpartisan analysis to the policy community.
    See examples of reports that are dependent on various AI definitions at \url{https://cset.georgetown.edu/reports}.}\
requires a solution for identifying AI-relevant research that is robust to the diversification of methods, tasks, and applications over time.
In this context, expert query development is increasingly impractical.
The solution that we describe in this paper embraces the dynamics of emerging technologies.

\section{Data}
\label{sec:data}

arXiv is organized into high-level domain repositories for physics, biology, computer science, statistics, and so forth.
Each of these repositories further defines a set of subjects to organize its content.
Authors select one or more subjects to describe each paper they submit.
Editors later review these subject tags \citep{clement2019use}.
arXiv's Computing Research Repository (CoRR) defines 39 subjects including artificial intelligence and machine learning.\footnote{~See \url{https://arxiv.org/category_taxonomy}.}

We focus in this paper on six subjects that CoRR editors describe as related to AI: Artificial Intelligence, Computation and Language (NLP), Computer Vision and Pattern Recognition (CV), Machine Learning,\footnote{~We include machine learning papers from the statistics repository (\texttt{stat.ML}) in this subject.
Cross-posting between the two categories is automatic.} Multiagent Systems, and Robotics.
According to CoRR documentation, the Artificial Intelligence subject ``[c]overs all areas of AI except Vision, Robotics, Machine Learning, Multiagent Systems, and Computation and Language (Natural Language Processing),'' because these areas have their own subjects.
It specifically ``includes Expert Systems, Theorem Proving [...], Knowledge Representation, Planning, and Uncertainty in AI.''
The Machine Learning subject ``[c]overs all aspects of machine learning research [and] is also an appropriate primary category for applications of machine learning methods.''
Because these applications may have their own subject areas, CoRR documentation specifies, ``If the primary domain of the application is available as another category in arXiv and readers of that category would be the main audience, that category should be primary.''
Some explicit examples of this are papers on CV, NLP, information retrieval, speech recognition, and neural networks.\footnote{\url{https://arxiv.org/corr/subjectclasses}.}

Using arXiv submissions in these categories as training data for subject classifiers, and defining AI-relevant research as the union of their positive predictions, is a useful framework for future researchers who may have differing needs or views on what constitutes AI.
Adding Neural and Evolutionary Computing or Information Retrieval papers might be warranted in future work.
We exclude them here for consistency with the CoRR editors' description of the Artifical Intelligence subject, but in practice, we suggest evaluating how sensitive quantities of interest are to these choices.

The compositional effect of including or excluding some subjects will be modest due to patterns of cross-posting papers across related subjects. 
There are 3,464 papers in our data with Information Retrieval as their primary subject, and 42\% also appear in one or more of the six subjects we consider AI-relevant here.
Of the 2,942 papers with the primary category of Neural and Evolutionary Computing, 39\% are cross-posted to at least one of our AI-relevant subjects, primarily Machine Learning.

From 2010 through 2019, authors submitted 1,060,321 papers to arXiv.\footnote{We restrict this effort to the last decade of arXiv papers to ensure reasonable numbers of papers in each subject in every year.} The largest repositories at the end of this decade, counting by papers' primary subjects, are physics (540,692), math (270,244), and computer science (194,627).
Table~\ref{tab:subject-counts} shows paper counts in the six computer science subjects we consider relevant.
There are 85,670 whose primary subject, the first selected by authors, is one of these six.
Authors can cross-post their papers under additional subjects, however, and when including these cross-posts there are 107,380 papers across the relevant subjects.

\begin{table}[h]
    \centering
    \caption{\
        arXiv contains 85,670 papers from 2010--2019 whose primary subject is one of the six we selected as relevant.
        107,380 papers, or an additional 21,710, appeared in at least one of the six subjects.
        This includes cross-posts from other subjects.}
    \begin{tabular}{lrr}
    \toprule
    Subject & Papers with Primary Subject & Papers Including Cross-posts  \\
    \midrule
      Artificial Intelligence (\texttt{cs.AI}) &    8,941 &   19,964 \\
      Natural Language Processing  (\texttt{cs.CL}) &   11,881 &   15,361 \\
      Computer Vision (\texttt{cs.CV}) &   28,309 &   35,254 \\
      Machine Learning (\texttt{cs.LG, stat.ML}) &   30,175 &   52,909 \\
      Multiagent Systems  (\texttt{cs.MA}) &     985 &    2,602 \\
      Robotics (\texttt{cs.RO}) &    5,379 &    7,933 \\
    \midrule
      Any of the above  &   85,670 &  107,380 \\
    \bottomrule
    \end{tabular}
    \label{tab:subject-counts}
\end{table}

Our targets for inference are larger corpora: Clarivate's Web of Science (WoS) Core Collection, Digital Science Dimensions, and Microsoft Academic Graph (MAG).
Training on arXiv is appealing for reasons we have described, but we ultimately care about performance in these more general knowledge bases, and many differences separate them.
The disciplinary coverage of the larger sources is broader, spanning fields in which we expect to find no AI-relevant papers.
For our analysis below, we create a combined corpus of unique English-language publications from Dimensions, MAG, and WoS in 2010 through 2019.
The result after deduplication is an analytic corpus of 38.6 million publications.\footnote{~We describe this process further in Appendix~\ref{appendix:corpus}.}

%MAG	11,365,414
%DS	2,753,459
%WOS	1,908,832
%DS ∩ MAG	5,000,325
%WOS ∩ MAG	1,203,339
%WOS ∩ DS	1,569,281
%WOS ∩ DS ∩ MAG	14,821,890

\section{Learning from arXiv}
\label{sec:methods}

From the arXiv corpus we draw two 10\% samples for development and testing, stratifying by publication year and subject label.
We use the resulting partition to train and evaluate solutions for identifying AI-relevant and subject-relevant publications.

Our baseline solution uses keyword matches.
We use 100 terms and patterns that we developed for a variety of document retrieval tasks in early Spring 2019, in a manual process: we reviewed search results and adapted the term list, and iterated until satsified.
(See Appendix \ref{appendix:keyword-list}.)
If one of these terms is present in the title or abstract of a publication, we consider that publication AI-relevant.
Our expectation was that this approach would achieve reasonable precision but low recall.
When tested against arXiv papers, considering papers in any of the six chosen subjects to be AI-relevant, we observe precision of .76 and recall of .43 ($F_1$ = .55).

A second approach for comparison is a keyword-classifier hybrid developed by Elsevier \citep{siebert_technical_2018} as part of a bibliometric study of AI.
The Elsevier group first extracted candidate terms from diverse textual sources, drawing from syllabi, books, patents, textbooks, the Cooperative Patent Classification scheme,\footnote{~\url{https://www.cooperativepatentclassification.org/cpcSchemeAndDefinitions}.} and AI news coverage.\footnote{~\url{https://aitopics.org}.}
The initial result was 800,000 keywords, which the group iteratively reduced to 797 distinct and specific terms.

The Elsevier team solicited comments on this set of terms from outside subject-matter experts.
Characteristically \citep{krafft_defining_2019}, however, these experts could not agree on any common set of keywords ``representative enough to scope the breadth of the field and [...] specific enough to AI'' \citep{siebert_technical_2018}.
The solution was for internal experts to score the terms on a three-point scale, and then task the outside experts with labeling a collection of publications that included the keywords.
This account illustrates the difficulty of delineating the field by consensus, and the investment that expert labeling entails.

Ultimately, incidence of the 797 terms in the input text was the basis for a series of features: variously weighted counts and proportions of lower- and higher-scoring terms in title and abstract text.
Following \citep{siebert_technical_2018}, we apply a random forest model to learn weights for these features using the training set drawn from the arXiv corpus.

We depart from a replication of the Elsevier method by training on arXiv, and the implementation details of doing so may not correspond with the original work.
Using a grid search to tune hyperparameter values and evaluating performance through cross-validation, we see precision of .74 and recall of .49 ($F_1$ = .59) in prediction of AI-relevant articles.
These results outperform our baseline keyword solution.\footnote{~For implementation details and replication code, see \url{https://github.com/georgetown-cset/ai-relevant-papers}.}

\begin{table}[h]
    \centering
    \caption{\
    Evaluation on arXiv test data shows $F_1$ scores of .84 for the all-subject SciBERT model and between .75 and .86 for subject-specific models.
    Our adaptation of the Elsevier AI model \citep{siebert_technical_2018} outperforms our keywords but falls behind the BERT models.}
    \begin{tabular}{lrrrrr}
    \toprule
    Method &  Precision &  Recall &  \quad $F_1$ \\
    \midrule
    CSET Keywords                                                       &       .76 &    .43 &            .55  \\
    Elsevier Keyword-classifier Hybrid \citep{siebert_technical_2018}    &       .74 &    .49 &            .59  \\
    SciBERT All Subjects                                               &\textbf{.83}&\textbf{.85}&\textbf{.84}\\
    SciBERT                                             &           &        &      \\
    \quad Natural Language Processing  (\texttt{cs.CL}) &       .86 &  .86  &  .86 \\
    \quad Computer Vision (\texttt{cs.CV})              &       .87 &  .81  &  .84 \\
    \quad Robotics (\texttt{cs.RO})                     &       .78 &  .73  &  .75 \\
    \bottomrule
    \end{tabular}
    \label{tab:test_performance}
\end{table}

Lastly, we apply SciBERT \citep{Beltagy2019Scibert}, a BERT \citep{devlin2018bert} model pre-trained on full text from Semantic Scholar then frozen and used to embed the title and abstract text of publications for classification.
Here we use the same tuning parameters as reported for the text classification task in \citep{Beltagy2019Scibert}.
We consider papers tagged with any of the six subjects to be AI-relevant and train a binary ``all subjects'' classifier.
In evaluation on the arXiv test set, we find improvements from SciBERT over the previous methods, with precision of .83 and recall of .85 ($F_1$ = .84).
We also train classifiers for AI-relevant subjects separately, one-versus-all.
This effort is successful for the three subjects that correspond with well-defined application fields: NLP ($F_1$ = .86), Computer Vision ($F_1$ = .84) and Robotics ($F_1$ = .75).

In Table~\ref{tab:test_performance}, we summarize the test performance of the baseline keyword solution, the Elsevier method, and the SciBERT models.
The all-subjects SciBERT model outperforms the alternative methods in the test data, and in comparison with the keyword-reliant solutions, we find appealing the availability of real-time updates from new arXiv content and the straightforward decomposability of AI-relevant research into subjects like computer vision.

\begin{figure}[h]
    \centering
    \caption{Higher performance from the supervised methods in more recent years is due in large part to longitudinal imbalance in the training data.
    Resampling or other strategies for imbalanced data can address this as appropriate for downstream analyses.
    The variation in keyword performance, by contrast, is the sign of a fast-moving field.}
    \includegraphics[width=.75\textwidth]{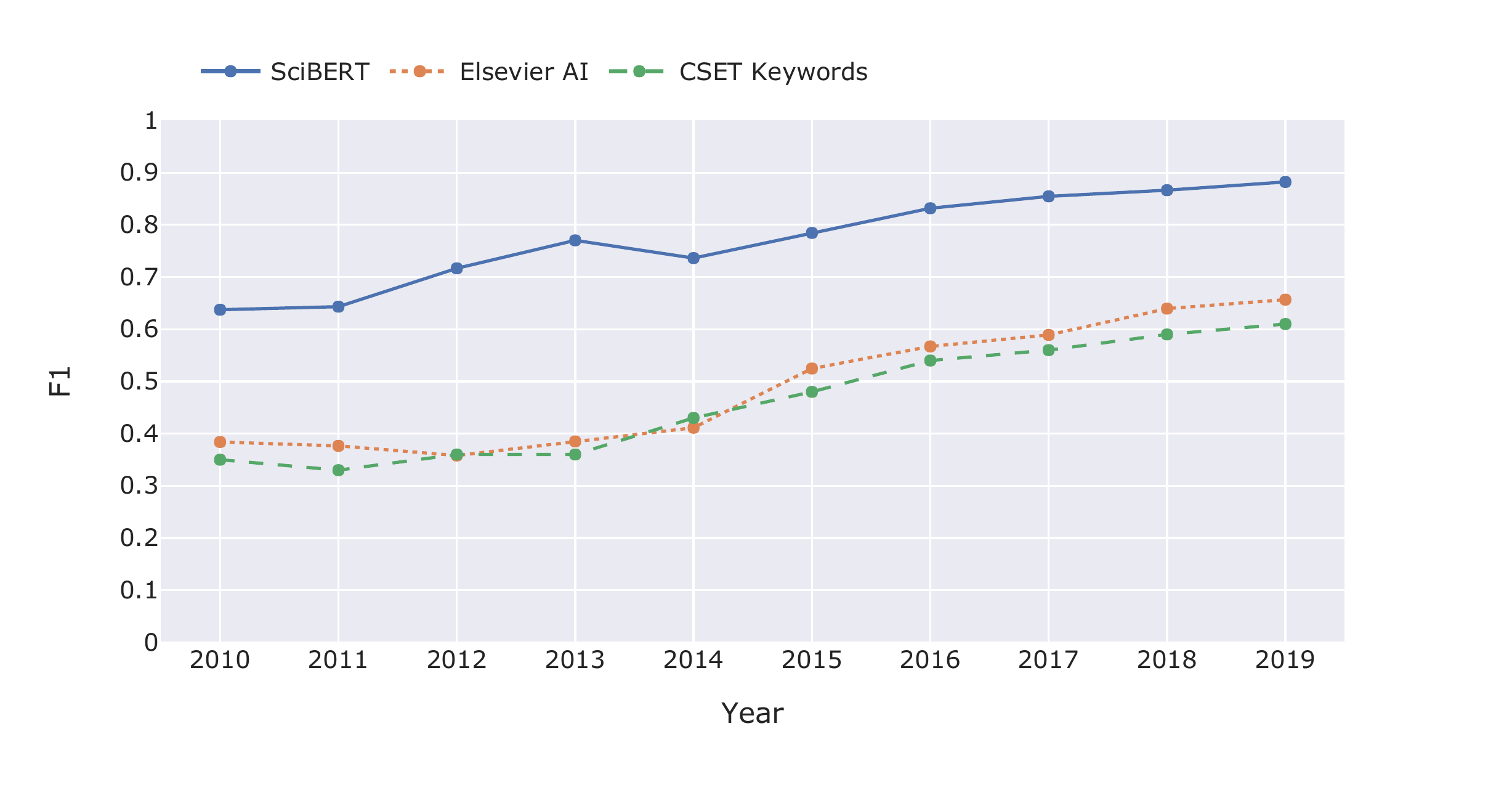}
    \label{fig:f1-by-year}
\end{figure}

In Figure \ref{fig:f1-by-year}, we assess the longitudinal performance of the keyword, Elsevier hybrid, and SciBERT classifiers.
The keyword solution performs best ($F_1$ = .61) in 2019, the year we developed it.
Its performance declines steadily in prior years, which we find unsurprising in a fast-moving field.
Elsevier's model and the SciBERT all-subjects model exhibit the same pattern, but for different reasons.

Higher performance from the supervised methods in recent years is due in large part to longitudinal imbalance in the training data.\footnote{\
    It is also possible that classification in earlier years is more difficult than in recent years, or for that matter easier, but the imbalance confounds direct evaluation.}
The appropriate response to this imbalance depends on the analytic context.
The expansion of arXiv since 2010 is attributable to its popularity relative to traditional journals, the growth of the particular fields arXiv covers, and secular trends in research output.
When training a classifier on arXiv for inference in WoS or elsewhere, one might seek the highest performance overall or prefer stable performance within strata meaningful in downstream analysis.
We suggest comparing the performance of a single model to that of period-specific models if inference focuses on time-series measures.

\section{Generalization}
\label{sec:generalization}

Because we lack gold labels for straightforward estimation of the models' performance outside of arXiv, we compare their predictions to other sources of subject information.
MAG provides a rich taxonomy of fields of study useful for this purpose.
Table~\ref{tab:fields} shows for top-level fields, along with subfields of computer science, the proportion of articles predicted relevant by each method.\footnote{\
    We necessarily restrict this table to publications found in MAG. These are 90\% of the unique articles across Dimensions, WoS, and MAG.}
The topical scope of MAG is broader than arXiv, so we approach generalization with some caution, limiting it to fields well-represented on arXiv.
During training, for example, the SciBERT classifiers encountered few papers in chemistry, medicine, or the social sciences.\footnote{\
We omit from prediction the MAG fields of Art, Business, Chemistry, Environmental science, Geography, History, Medicine, Philosophy, Political science, Psychology, and Sociology.}

Table~\ref{tab:fields} shows for top-level fields, along with subfields of computer science, the proportion of articles predicted relevant by each method.
Each row in the table represents publications in a MAG field, and each column a method or model.
``SciBERT'' refers to the All Subjects model, and from left to right, the arXiv subject abbreviations refer to the Computation and Language (NLP), Computer Vision, and Robotics subject models.

Plausibly, the keyword, Elsevier, and SciBERT methods for identifying AI-relevant publications yield the highest prediction rates in artificial intelligence, computer vision, data mining, machine learning, natural language processing, pattern recognition, and speech recognition.
Consistent with test performance, which showed higher recall for the all-subject SciBERT model (.85) than the hybrid (.49) or keyword (.43) methods, the SciBERT model tends to predict much larger proportions of these fields to be relevant.
The MAG fields of study are themselves estimates, however, so this is a validation exercise rather than an evaluation against ground truth.\footnote{\
    MAG provides field scores for each paper: the positive subset of cosine similarities between its embedding and those of fields.
    Here we consider a paper to belong in a field of study if its score is positive.
}

The final columns of Table~\ref{tab:fields} give corresponding statistics for the subject-specific SciBERT models.
The NLP (\texttt{cs.CL}) model identifies 77\% of papers in MAG's natural language processing field as relevant, along with 22\% of the speech recognition field and 18\% of information retrieval papers.
The subject model successfully discriminates between NLP papers and those in machine learning (only 7\% relevant) or artificial intelligence (8\%).\footnote{\
    Like arXiv subjects, MAG fields are non-exclusive.
    Many papers have positive field scores for more than one field.
}
Predictions from the computer vision (\texttt{cs.CV}) model identify 53\% of the computer vision field and 54\% of pattern recognition papers as relevant.
Positive predictions from the robotics (\texttt{cs.RO}) model are relatively rare, but it identifies 71\% of papers in the robotics subfield of engineering and mathematics as relevant, along with 17\% of the simulation subfield and 11\% of human-computer interaction.

%Some rates of predicted relevance in MAG fields outside of computer science point to likely concentrations of false positives.
%In philosophy, the NLP (\texttt{cs.CL}) SciBERT model identifies 10\% of papers as relevant, the all-subject model 6\%, and the Elsevier model 3\%.
%We see a 2\% positive-prediction rate from the all-subject model in psychology, where the Elsevier model also identifies 4\% of papers as relevant.

\parnoteclear
\begin{table}[h]
    \centering
    \caption{\
    Each row describes a MAG field of study.
    Column ``Count'' reports the number of publications in the field, and the remaining columns give the percent of the field identified as relevant by each model or method.
    For example, the SciBERT all-subjects model identifies 66\% of MAG's 1.2M CS / Artificial Intelligence publications as relevant, from its training on arXiv papers.
    For fields that correspond with AI-relevant arXiv subjects, the highest percentage in a row is bold.
    From left to right, the arXiv subject model abbreviations  refer to for Computation and Language (NLP), Computer Vision, and Robotics.
    We observe plausible rates of predicted relevance from SciBERT models across MAG fields.
    }
    {\small
    \newcolumntype{R}[1]{>{\raggedleft\arraybackslash}p{#1}}
    \begin{tabularx}{\textwidth}{l r *6{R{1.2 cm}}}
        \toprule
        {} & &  \multicolumn{6}{l}{Percent of Count predicted relevant}
        \tabularnewline \cmidrule(lr){3-8}
        {} & &  & & SciBERT & SciBERT& SciBERT& SciBERT   \\
        MAG Field / Subfield &     Count        &  Keywords         & Elsevier & All Subj. &  \texttt{cs.CL} &  \texttt{cs.CV} &  \texttt{cs.RO} \\
        \midrule
        Biology                           &   8,820,224 &       1 &         1 &        1 &           0 &           0 &           1 \\
        CS / Algorithm                    &    403,571 &       14 &        17 &       26 &           1 &           8 &           2 \\
        CS / Artificial intelligence       &   1,243,775 &      39 &        38 &       66 &           8 &          31 &           6 \\
        CS / Computational science        &     18,629 &        5 &         5 &        5 &           0 &           1 &           1 \\
        CS / Computer architecture        &     15,018 &       11 &        11 &        7 &           0 &           1 &           1 \\
        CS / Computer engineering         &     20,994 &       15 &        16 &       14 &           0 &           4 &           2 \\
        CS / Computer graphics (images)   &     58,976 &       10 &         5 &       30 &           0 &          23 &           3 \\
        CS / Computer hardware            &    115,751 &        5 &         3 &        6 &           0 &           2 &           2 \\
        CS / Computer network             &    418,390 &        3 &         5 &        2 &           0 &           0 &           0 \\
        CS / Computer security            &    220,493 &        5 &         5 &        5 &           0 &           1 &           1 \\
        CS / Computer vision              &    494,902 &       29 &        23 & 64 &        0 &          53 &           9 \\
        CS / Data mining                  &    345,223 &       28 &        31 &       42 &           4 &           6 &           1 \\
        CS / Data science                 &    105,878 &       14 &        17 &       17 &           4 &           1 &           0 \\
        CS / Database                     &    102,016 &        6 &         7 &        9 &           1 &           2 &           1 \\
        CS / Distributed computing        &    276,100 &        7 &        12 &        9 &           0 &           1 &           2 \\
        CS / Embedded system              &    125,784 &        4 &         4 &        8 &           0 &           1 &           4 \\
        CS / Human–computer interaction   &    129,101 &       11 &        15 &       31 &           2 &           3 &          11 \\
        CS / Information retrieval        &    108,145 &       28 &        27 &       44 &          18 &           6 &           0 \\
        CS / Internet privacy             &     80,802 &        2 &         3 &        2 &           1 &           0 &           0 \\
        CS / Knowledge management         &    318,313 &        3 &         8 &        6 &           1 &           0 &           0 \\
        CS / Library science              &    166,741 &        1 &         1 &        1 &           1 &           0 &           0 \\
        CS / Machine learning             &    360,586 &       51 &        55 &       71 &           7 &          15 &           2 \\
        CS / Multimedia                   &    219,419 &        6 &         9 &       11 &           2 &           2 &           1 \\
        CS / Natural language processing  &    103,318 &       38 &        41 &       79 &          77 &           5 &           0 \\
        CS / Operating system             &     50,324 &        2 &         2 &        2 &           0 &           0 &           1 \\
        CS / Parallel computing           &     77,951 &        7 &         8 &        6 &           0 &           2 &           0 \\
        CS / Pattern recognition          &    360,826 &       52 &        48 &      80 &           3 &          54 &           1 \\
        CS / Programming language         &     50,998 &        5 &        10 &        9 &           3 &           0 &           1 \\
        CS / Real-time computing          &    271,586 &        8 &        10 &       12 &           0 &           3 &           3 \\
        CS / Simulation                   &    280,108 &        6 &         9 &       23 &           0 &           2 &          17 \\
        CS / Software engineering         &     77,539 &        4 &        10 &        8 &           1 &           0 &           2 \\
        CS / Speech recognition           &    106,362 &       41 &        37 &       58 &          22 &          11 &           1 \\
        CS / Telecommunications           &     86,710 &        1 &         2 &        1 &           0 &           0 &           0 \\
        CS / Theoretical computer science &    152,733 &       11 &        16 &       20 &           1 &           2 &           1 \\
        CS / World Wide Web               &    228,179 &        6 &         9 &        8 &           3 &           0 &           0 \\
        Economics                         &   3,370,477 &       1 &         2 &        1 &           0 &           0 &           0 \\
        Engineering                       &   6,518,254 &       3 &         5 &        6 &           0 &           0 &           4 \\
        Engineering / Robotics            &      29,488 &      21 &        25 & 72 &           0 &           9 &          71 \\
        Geology                           &   1,610,737 &       2 &         2 &        3 &           1 &           1 &           2 \\
        Materials science                 &   2,407,580 &       0 &         1 &        1 &           0 &           0 &           1 \\
        Mathematics                       &   4,032,139 &       5 &         8 &        8 &           0 &           1 &           2 \\
        Physics                           &   4,195,403 &       1 &         1 &        1 &           0 &           0 &           0 \\
        \bottomrule
    \end{tabularx}
    \parnotes
    }
    \label{tab:fields}
\end{table}

\section{Conclusion}
\label{sec:conclusion}

Our results demonstrate high classification performance from SciBERT \citep{Beltagy2019Scibert} models applied to learning arXiv subjects.
Although we did not evaluate SciBERT against a comparable BERT model pre-trained on Wikipedia and the BookCorpus \citep{devlin2018bert}, we attribute some of this performance to transfer learning via SciBERT's embedding of scientific vocabulary after pre-training on Semantic Scholar.
Within the set of topics the models saw in training on arXiv papers, inference in WoS appears feasible: we observe plausible rates of predicted relevance in MAG fields of study.

Looking forward, manual annotation is the obvious solution to our lack of labeled examples in Dimensions, MAG, and WoS.
However, developing guidelines for labeling publications for AI-relevance would require addressing definitional questions we sidestepped in this work; it would represent a departure from using the implicit delineation of the field provided by arXiv preprints.
But we anticipate that labeling examples to approximate the boundaries of arXiv subjects, like NLP and computer vision, is far more tractable than manual labeling for AI relevance.

The arXiv corpus exhibits a class imbalance of about 9:1 in favor of negative examples.
In the analytic corpus, whose topical coverage is broader, we assume the true imbalance is greater.
The appropriate tuning for class performance will depend on the application.

Another major direction for future work is expanding domain generalizibility, particularly in potential application areas.
We have substantive interest in papers on topics unavailable in arXiv, from agriculture to medicine.
We would consider reports of AI applications in trade journals to be AI-relevant in principle, for example, but we focus in this paper on a delineation of the field whose implementation may not include them.
To expand into these areas, we anticipate leveraging bibliometric data in addition to text: applying scientometric methods to extend the identification of publications describing the development and applications of AI beyond arXiv's coverage.

\bibliography{references}

\newpage
\appendix
\section{Further results}
\setcounter{table}{0}
\renewcommand{\thetable}{\Alph{section}.\arabic{table}}

Table~\ref{tab:keywords_by_year} reports the evaluation of keywords (Appendix ~\ref{appendix:keyword-list}) in the full arXiv data by year.
Scores are for the positive class.
The ``Support'' column refers to the number of AI-relevant articles out of the ``Total`` articles, where AI-relevance is defined as elsewhere by having at least one of the six selected subject tags: \texttt{cs.AI}, \texttt{cs.CL}, \texttt{cs.CV}, \texttt{cs.LG}/\texttt{stat.ML}, \texttt{cs.MA}, and \texttt{cs.RO}.
Performance in highest in 2019, when we generated the terms.
We take the declining performance in earlier years to suggest the need for continuous maintenance of keywords.
%TODO: this is the place to discuss why recall declines, to Beth's point

\begin{table}[h]
    \centering
    \caption{Keyword performance in full arXiv data.}
    \begin{tabular}{lrrrrr}
    \toprule
Year &  Precision &  Recall &   $F_1$ & Support  &   Total \\
\midrule                                                
2010 &       .50 &    .27 &  .35 &  1,379 &   70,286 \\
2011 &       .54 &    .24 &  .33 &  2,025 &   76,605 \\
2012 &       .63 &    .25 &  .36 &  3,370 &   84,389 \\
2013 &       .65 &    .25 &  .36 &  4,561 &   92,866 \\
2014 &       .66 &    .31 &  .43 &  4,896 &   97,598 \\
2015 &       .71 &    .36 &  .48 &  6,663 &  105,128 \\
2016 &       .78 &    .41 &  .54 & 10,566 &  113,436 \\
2017 &       .77 &    .44 &  .56 & 15,670 &  123,781 \\
2018 &       .77 &    .48 &  .59 & 23,891 &  140,392 \\
2019 &       .80 &    .49 &  .61 & 34,359 &  155,840 \\
\midrule                                                
All &       .76 &    .43 &  .55 &  103,380 &  1,060,321 \\
    \bottomrule
    \end{tabular}
    \label{tab:keywords_by_year}
\end{table}

In Table~\ref{tab:elsevier-performance}, we show the test performance of the keyword-classifier hybrid developed by Elsevier.
This solution shows improvements over our baseline keyword solution.
We attribute higher performance in more recent years to longitudinal imbalance in the training data.
There is also a class imbalance of about 9:1 in favor of negative examples.
Its effect on performance is apparent despite the use of class weights.

\begin{longtable}[c]{lrrrrrrrrrr}
\caption{Elsevier keyword-classifier performance in arXiv test data.} \\
\toprule
{}         &	\multicolumn{4}{l}{\textit{Positive Class}}         & \multicolumn{4}{l}{\textit{Negative Class}}  & \multicolumn{2}{l}{\textit{Wtd. Avg.}} \\
Year       &  Precision & Recall & $F_1$ &  Support  &  Precision &  Recall &  $F_1$    &  Support & $F_1$  & Support \\  
\midrule
2010 &      .50 &   .31 & .38 &     138 &      .99 &   .99 & .99 &   6,891 &      .98 &    7,029 \\
2011 &      .50 &   .30 & .38 &     202 &      .98 &   .99 & .99 &   7,458 &      .97 &    7,660 \\
2012 &      .58 &   .26 & .36 &     337 &      .97 &   .99 & .98 &   8,102 &      .96 &    8,439 \\
2013 &      .60 &   .28 & .39 &     456 &      .96 &   .99 & .98 &   8,831 &      .95 &    9,287 \\
2014 &      .59 &   .31 & .41 &     489 &      .96 &   .99 & .98 &   9,271 &      .95 &    9,760 \\
2015 &      .69 &   .42 & .52 &     666 &      .96 &   .99 & .97 &   9,847 &      .95 &   10,513 \\
2016 &      .75 &   .45 & .57 &   1,057 &      .95 &   .98 & .97 &  10,287 &      .93 &   11,344 \\
2017 &      .74 &   .49 & .59 &   1,567 &      .93 &   .98 & .95 &  10,811 &      .91 &   12,378 \\
2018 &      .75 &   .55 & .64 &   2,389 &      .91 &   .96 & .94 &  11,650 &      .89 &   14,039 \\
2019 &      .81 &   .55 & .66 &   3,436 &      .88 &   .96 & .92 &  12,148 &      .86 &   15,584 \\
\midrule
All  &      .74 &   .49 & .59 &  10,737 &      .94 &   .98 & .96 &  95,296 &      .92 &  106,033 \\
\bottomrule
\label{tab:elsevier-performance}
\end{longtable}

Table~\ref{tab:scibert-any-ai} gives test performance of the all-subject SciBERT model.
Like the Elsevier solution, the best results are for recent years, due to longitudinal imbalance.

\newpage
\section{Keywords}
\label{appendix:keyword-list}
\setcounter{table}{0}
\renewcommand{\thetable}{\Alph{section}.\arabic{table}}
{\setlength{\tabcolsep}{12pt}
    \begin{table}[h!]
        \centering
        \caption{We use these terms and patterns in our baseline search strategy.
        Originally, we developed this list for document retrieval tasks on a variety of knowledgebases, such as WoS, ProQuest, Dimensions, and CNKI, in early Spring 2019.
        The * character represents a wildcard that matches zero or more non-whitespace characters.}
        \begin{tabular}{ll}
            \toprule
            active learning	&	incremental clustering	\\
            adaptive learning	&	information extraction	\\
            anomaly detection	&	information fusion	\\
            artificial intelligence	&	information retrieval	\\
            associative learning	&	k nearest neighbor	\\
            autonomous navigation	&	knowledge based system*	\\
            autonomous system*	&	knowledge discovery	\\
            autonomous vehicle*	&	knowledge representation	\\
            average link clustering	&	language identification	\\
            back propagation	&	machine learning	\\
            backpropagation	&	machine perception	\\
            binary classification	&	machine translation	\\
            bioNLP	&	multi class classification	\\
            boltzmann machine	&	multi label classification	\\
            character recognition	&	multi task learning	\\
            classification algorithm	&	natural language generation	\\
            classification label*	&	natural language processing	\\
            clustering method*	&	natural language understanding	\\
            complete link clustering	&	neural network	\\
            computer aided diagnosis	&	object recognition	\\
            computer vision	&	one shot learning	\\
            deep learning	&	pattern matching	\\
            ensemble learning	&	pattern recognition	\\
            evolutionary algorithm	&	random forest	\\
            fac* expression recognition	&	recommend* system*	\\
            fac* identification	&	recurrent network	\\
            fac* recognition	&	reinforcement learning	\\
            feature extraction	&	scene* classification	\\
            feature learning	&	scene* understanding	\\
            feature matching	&	self driving car*	\\
            feature selection	&	semi supervised learning	\\
            feature vector	&	sentiment classification	\\
            feedforward network	&	single link clustering	\\
            feedforward neural network	&	spatial learning	\\
            fuzzy clustering	&	speech processing	\\
            generative adversarial network	&	speech recognition	\\
            gradient algorithm	&	speech synthesis	\\
            graph matching	&	statistical learning	\\
            graphical model	&	strong artificial intelligence	\\
            handwriting recognition	&	supervised learning	\\
            hierarchical clustering	&	support vector machine	\\
            hierarchical model	&	text mining	\\
            human robot	&	text processing	\\
            image annotation	&	transfer learning	\\
            image classification	&	translation system	\\
            image matching	&	unsupervised learning	\\
            image processing	&	video classification	\\
            image registration	&	video processing	\\
            image representation	&	weak artificial intelligence	\\
            image retrieval	&	zero shot learning	\\
            \bottomrule
        \end{tabular}
        \label{tab:keywords_by_subject}
    \end{table}}

\newpage
\begin{longtable}[c]{lrrrrrrrrrr}
\caption{All-subject SciBERT performance in arXiv test data.}\\
\toprule
{}         &	\multicolumn{4}{l}{\textit{Positive Class}}         & \multicolumn{4}{l}{\textit{Negative Class}}  & \multicolumn{2}{l}{\textit{Wtd. Avg.}} \\
Year       &  Precision & Recall & $F_1$ &  Support  &  Precision &  Recall &  $F_1$    &  Support & $F_1$  & Support \\  
\midrule                                                                                                       
	2010    &   .64 &    .63 &      .64 &     138   &     .99 &     .99 &       .99 &     6,891     &          .99  &        7,029            \\   
	2011    &   .65 &    .63 &      .64 &     202   &     .99 &     .99 &       .99 &     7,458     &          .98  &        7,660            \\   
	2012    &   .74 &    .69 &      .72 &     337   &     .99 &     .99 &       .99 &     8,102     &          .98  &        8,439            \\   
	2013    &   .80 &    .74 &      .77 &     456   &     .99 &     .99 &       .99 &     8,831     &          .98  &        9,287            \\   
	2014    &   .74 &    .73 &      .74 &     489   &     .99 &     .99 &       .99 &     9,271     &          .97  &        9,760            \\   
	2015    &   .78 &    .79 &      .78 &     666   &     .99 &     .98 &       .99 &     9,847     &          .97  &       10,513            \\   
	2016    &   .82 &    .84 &      .83 &    1,057   &     .98 &     .98 &       .98 &    10,287    &          .97  &       11,344             \\   
	2017    &   .83 &    .89 &      .85 &    1,567   &     .98 &     .97 &       .98 &    10,811    &          .96  &       12,378             \\   
	2018    &   .83 &    .90 &      .87 &    2,389   &     .98 &     .96 &       .97 &    11,650    &          .95  &       14,039             \\   
	2019    &   .87 &    .89 &      .88 &    3,436   &     .97 &     .96 &       .97 &    12,148    &          .95  &       15,584             \\   
\midrule                                                                                                                       
	All &   .83 &    .85 &      .84 &   10,737   &     .98 &     .98 &       .98 &    95,296   &          .97 &      106,033 \\
\bottomrule
\label{tab:scibert-any-ai}
\end{longtable}

\section{Analytic corpus}
\label{appendix:corpus}
\setcounter{table}{0}

To create the analytic corpus we linked or deduplicated the contents of Dimensions, MAG (excluding datasets and patents), and WoS.
After normalizing titles, abstracts, and author last names, we considered records within or across datasets to represent the same publication if they shared at least three non-null values across title, abstract, publication year, author surnames,
citations (for within-dataset matches), and DOI.
We restricted the result to English-language publications since 2010 with non-null titles and abstracts.
For language detection we used CLD2, as implemented in \href{https://pypi.org/project/pycld2/}{PYCLD2}.
\end{document}